%% file: main.tex
\documentclass[sigconf]{acmart}

\usepackage{amsfonts,amssymb}
\usepackage{amsthm} % 如果要使用proof语句就必须要引入这个包
\usepackage{multirow}
\usepackage{bbding}
\usepackage{makecell}
\usepackage{float}
\usepackage{enumitem}
\usepackage{overpic}
\usepackage{graphicx}
\usepackage{subfigure}
\usepackage{wasysym}
\usepackage{utfsym}
\usepackage{hyperref}
\PassOptionsToPackage{table,xcdraw}{xcolor}

\AtBeginDocument{%
}

\copyrightyear{2024}
\acmYear{2024}
\setcopyright{rightsretained}
\acmConference[SIGIR '24]{Proceedings of the 47th International ACM SIGIR Conference on Research and Development in Information Retrieval}{July 14--18, 2024}{Washington, DC, USA}
\acmBooktitle{Proceedings of the 47th International ACM SIGIR Conference on Research and Development in Information Retrieval (SIGIR '24), July 14--18, 2024, Washington, DC, USA}
\acmDOI{10.1145/3626772.3657669}
\acmISBN{979-8-4007-0431-4/24/07}


\begin{document}

\input{Headings/head.tex}
\input{Headings/abstract.tex}
\input{Headings/keywords.tex}
%%
%% This command processes the author and affiliation and title
%% information and builds the first part of the formatted document.
\maketitle

\input{Sections/1.introduction}
\input{Sections/2.related_work}

\input{Sections/3.methodology}
\input{Sections/4.applications}
\input{Sections/5.case_example}
\input{Sections/6.conclusion}

\newpage
%%
%% The next two lines define the bibliography style to be used, and
%% the bibliography file.
\bibliographystyle{ACM-Reference-Format}
\balance
\bibliography{reference}

% \newpage
% \newpage
\appendix

\input{Appendix/appendix.tex}
\input{Appendix/acknowledgment.tex}

\end{document}

%% file: Headings/head.tex
%%
%% The "title" command has an optional parameter,
%% allowing the author to define a "short title" to be used in page headers.
% \title{Pretrained Language Models based Recommendation Intent Identification in Conversation}
% \title{Recommendation Intent Identification in Conversations with Prompt Learning}
% \title{Recommendation Intent Identification in Conversations with Pretrained Language Models}
\title{MACRec: a Multi-Agent Collaboration Framework for Recommendation}
% \title{Multi-Agent Collaboration Framework for Recommender Systems}

%%
%% The "author" command and its associated commands are used to define
%% the authors and their affiliations.
%% Of note is the shared affiliation of the first two authors, and the
%% "authornote" and "authornotemark" commands
%% used to denote shared contribution to the research.
% \author{}
% \email{}
% \affiliation{
% }

\author{Zhefan Wang}
\authornote{Both authors contributed equally to this research.}
\affiliation{
  \institution{DCST, Tsinghua University}
  \city{Beijing 100084}
  \country{China}
  \postcode{100084}
}
\email{wzf23@mails.tsinghua.edu.cn}

\author{Yuanqing Yu}
\authornotemark[1]
\affiliation{
  \institution{DCST, Tsinghua University}
  \city{Beijing 100084}
  \country{China}
  \postcode{100084}
}
\email{yyq23@mails.tsinghua.edu.cn}

\author{Wendi Zheng}
\affiliation{
  \institution{DCST, Tsinghua University}
  \city{Beijing 100084}
  \country{China}
  \postcode{100084}
}
\email{zhengwd23@mails.tsinghua.edu.cn}

\author{Weizhi Ma}
\authornote{Corresponding author. This work is supported by the Natural Science Foundation of China (Grant No. U21B2026, 62372260).}
\affiliation{%
  \institution{AIR, Tsinghua University}
  \city{Beijing 100084}
  \country{China}
  \postcode{100084}
}
\email{mawz@tsinghua.edu.cn}

\author{Min Zhang}
\authornotemark[2]
\affiliation{%
  \institution{DCST, Tsinghua University}
  \city{Beijing 100084}
  \country{China}
  \postcode{100084}
}
\email{z-m@tsinghua.edu.cn}

%%
%% By default, the full list of authors will be used in the page
%% headers. Often, this list is too long, and will overlap
%% other information printed in the page headers. This command allows
%% the author to define a more concise list
%% of authors' names for this purpose.
\renewcommand{\shortauthors}{Zhefan Wang, Yuanqing Yu, Wendi Zheng, Weizhi Ma, \& Min Zhang}
%% No italics


\ccsdesc[500]{Information systems~Recommender systems}

%% file: Headings/abstract.tex
%%
%% The abstract is a short summary of the work to be presented in the
%% article.
\begin{abstract}
% Agents 因其决策能力和完成复杂任务备受关注
LLM-based agents have gained considerable attention for their decision-making skills and ability to handle complex tasks.
% Motivation: 考虑到目前基于agents进行推荐的工作在多智能体合作的探索不足，且缺少开源框架，我们提出了MACRec框架。
Recognizing the current gap in leveraging agent capabilities for multi-agent collaboration in recommendation systems, we introduce \textbf{MACRec}, a novel framework designed to enhance recommendation systems through multi-agent collaboration.
% 具体介绍
Unlike existing work on using agents for user/item simulation, we aim to deploy multi-agents to tackle recommendation tasks directly.
In our framework, recommendation tasks are addressed through the collaborative efforts of various specialized agents, including \textsl{Manager}, \textsl{User/Item Analyst}, \textsl{Reflector}, \textsl{Searcher}, and \textsl{Task Interpreter}, with different working flows.
Furthermore, we provide application examples of how developers can easily use MACRec on various recommendation tasks, including rating prediction, sequential recommendation, conversational recommendation,
% interactive recommendation, click-through rates (CTR) prediction, 
and explanation generation of recommendation results.
% 开源 % The framework is available at https://github.com/wzf2000/MACRec.
% The framework and its documentation and examples are made publicly available at https://github.com/wzf2000/MACRec.
The framework and demonstration video are publicly available at \href{https://github.com/wzf2000/MACRec}{https://github.com/wzf2000/MACRec}.

\end{abstract}

% LLM-based agents have gained considerable attention for their decision-making skills and ability to handle complex tasks. There are various complex decision-making tasks in recommendation scenarios, so some previous studies have tried to use agents in this scenario. However, existing work focuses on using agents for user/item simulation, ignoring the multi-agent collaboration ability for recommendation. In this paper, we introduce MACRec, a novel framework designed to enhance recommendation systems through multi-agent collaboration, to deploy multi-agents for various recommendation tasks directly. In our framework, recommendation tasks are addressed through the collaborative efforts of various specialized agents, including Manager, User/Item Analyst, Reflector, Searcher, and Task Interpreter, with different working flows. Furthermore, we provide application examples of how developers can easily use MACRec on various recommendation tasks, including rating prediction, sequential recommendation, conversational recommendation, and explanation generation of recommendation results. The framework and demonstration video are publicly available at https://github.com/wzf2000/MACRec.

%% file: Headings/keywords.tex
%%
%% The code below is generated by the tool at http://dl.acm.org/ccs.cfm.
%% Please copy and paste the code instead of the example below.
%%
%%
%% Keywords. The author(s) should pick words that accurately describe
%% the work being presented. Separate the keywords with commas.

\keywords{Multi-agents; Large Language Models; Recommender Systems}
% \keywords{Multi-agents}

%% file: Sections/1.introduction.tex
\section{Introduction}
% 1. 研究背景: 推荐系统 & LLM-based Agents
% % 推荐系统在日常生活中的重要作用。
Recommender systems (RSs) play a vital role in improving user experience and platform economic benefits, which have become an essential part of various domains, such as e-commerce, social media, and so on. 
% % 随着LLM的发展，LLM-based agents在做决策和完成复杂任务方面展现出优秀能力。这些模型的语义理解、规划和决策能力为推荐系统领域提供了新的可能性。
Currently, the advancement of Large Language Models (LLMs)~\cite{brown2020language, gpt42023technical, touvron2023llama, zeng2022glm} has introduced LLM-based agents~\cite{yao2022react, nakano2021webgpt, shen2023hugginggpt} capable of completing complex tasks. These agents' semantic understanding, planning, and decision-making skills unlock new potentials for more nuanced and context-aware recommendations.
% With the recent advances in Large Language Models (LLMs)~\cite{brown2020language, gpt42023technical}, LLM-based agents such as ReAct~\cite{yao2022react}, WebGPT~\cite{nakano2021webgpt}, and HuggingGPT~\cite{shen2023hugginggpt} have shown exceptional capabilities in completing complex tasks. The verbal comprehension, planning, and decision-making abilities of these models offer new opportunities to enrich recommendations.

% 2. 研究现状（相关工作）
% % 研究者们已经开始利用agents的能力解决推荐任务。
Researchers have started to utilize the capabilities of agents to solve recommendation tasks. 
% % 1）一大类工作例如RecAgent，AgentCF，主要侧重于使用agents来模拟用户或物品行为，提供对用户偏好的见解，但无法集成到RS中。
Existing work like ~\cite{zhang2023generative, wang2023RecAgent, zhang2023agentcf} primarily focuses on employing agents for simulating user or item behaviors, providing insights into user preferences but falling short of integration into RSs.
% % 2）一些工作尝试利用代理的能力直接构建recommender，主要使用单一类型的agent, 带有planning和memory组件以及辅助工具（如搜索引擎）。
On the other hand, some studies~\cite{wang2023recmind, huang2023InteRecAgent} attempt to leverage the capabilities of agents to directly build a recommender, primarily using one single agent with planning and memory components and auxiliary tools (e.g., search engine).
% % % 然而，
% Motivation：推荐决策很复杂，单个agent不能很好地应对，缺乏对多智能体的尝试（有用且重要）
However, there are various complex decision-making tasks in recommendation scenarios~\cite{sun2023neighborhood,sun2024collaborative}, on which single-agent instances are unable to perform well.
% It has been demonstrated that single-agent instance performs more poorly than multi-agent collaboration in reasoning, factuality, and question-answering tasks~\cite{liang2023encouraging, du2023improving}.
Multi-agent collaboration, which is near to human workflows, is believed to accomplish complex tasks better with collective intelligence.
Although work~\cite{shu2023rah} proposes a multi-agent recommendation framework, it only has limited agent types and a fixed collaboration mode.

\begin{table*}[t!]
\centering
\caption{Comparison between previous work and our MACRec. Note that \textsl{Single-type Agents} indicate all agents serve the same role (e.g., users), while \textsl{Multi-type Agents} refer to agents having multiple roles and capabilities (e.g., managers, reflectors).}
\label{tab:related_work}
\begin{tabular}{cccccc}
\toprule

\textbf{Model} & \textbf{Objectives} & \textbf{Single-type Agents} & \textbf{Multi-type Agents}  & \textbf{Diverse Rec. Scenarios} & \textbf{Open-source} \\
\midrule
RecAgent~\cite{wang2023RecAgent} & User Simulation & \Checkmark &  &  & \Checkmark \\
Agent4Rec~\cite{zhang2023generative} & User  Simulation & \Checkmark &  &  &  \Checkmark \\
AgentCF~\cite{zhang2023agentcf} & U-I Inter Simulation & & \Checkmark &  &  \\
\midrule
RAH~\cite{shu2023rah} & Recommender &  & \Checkmark &  &  \\
RecMind~\cite{wang2023recmind}  & Recommender & \Checkmark &  & \Checkmark &  \\
InteRecAgent~\cite{huang2023InteRecAgent}  & Recommender & \Checkmark &  &  & \\
\midrule
\textbf{MACRec} & Recommender  & \Checkmark & \Checkmark & \Checkmark & \Checkmark \\
\bottomrule
\end{tabular}
\end{table*}

% 3. Our Framework
To better unleash the potential of multi-agent collaboration for recommendation tasks, we propose \textbf{MACRec}, a novel \textbf{M}ulti-\textbf{A}gent \textbf{C}ollaboration framework for recommender systems, designed to harness the diverse capabilities of each agent. 
Notably, this framework differs from studies for simulation with agents but focuses on building a recommender directly.
MACRec provides customizable agents with abilities powered by LLMs and useful tools. For example, we offer \textsl{Manager} to plan and manage task execution, \textsl{Reflector} to reflect on previous errors, \textsl{User/Item Analysts} to analyze user/item characteristics, \textsl{Searcher} to search more information using the search tool, and \textsl{Task Interpreter} to translate the dialogs into executable recommendation tasks.
These agents with different roles work collaboratively to tackle a specific recommendation task.

% Applications
Additionally, we provide application examples to use MACRec on various recommendation tasks, including rating prediction, sequential recommendation, conversational recommendation,
% interactive recommendation, click-through rates (CTR) prediction,
and explanation generation of recommendation results. 
Considering the varying requirements for agents in different scenarios, we showcase examples of selecting and customizing agents to collaborate on diverse recommendation tasks.
% Understandably, different recommendation scenarios require different sets of agents.
% Our case experiments on rating prediction and sequential recommendation tasks show our framework's practical feasibility and effectiveness. 
Furthermore, we developed an online web interface for our MACRec, providing a user-friendly visualization of the agents' collaboration process.
The main strengths of this work can be summarized as follows:
\begin{itemize}[leftmargin=*]
    \item \textbf{A New Multi-agent Collaboration Framework for Recommendation.}
    % 不同于以往的simulation工作，提出一个新的多智能体合作框架。提供可定制的
    Unlike previous studies focused on user/item simulation with agents, we propose a new multi-agent collaboration framework for recommendation \textbf{MACRec}. In this framework, agents with different abilities, work collaboratively are involved to tackle specific recommendation tasks.
    % \item \textbf{Flexible Collaboration Modes.}
    % \item \textbf{Adaptive Configuration for Distinct Tasks.}
    % support the adaptive customization of agents and how they collaborate across different tasks.
    % The agents in our framework can achieve different collaboration modes by customizing the prompts and system settings.
    
    % By customizing settings for data and agents, the framework can adaptively support numerous tasks.
    % \item \textbf{Adaptive Configuration for various tasks.}
    
    \item \textbf{Diverse Applications on Recommendation Scenarios.}
    We present application examples on various recommendation scenarios, including rating prediction, sequential recommendation, explanation generation,
    % interactive recommendation, CTR prediction, 
    and conversational recommendation.

    \item \textbf{A User-friendly Online Web Interface. } We developed an online web interface for MACRec, visualizing how agents collaboratively tackle tasks.

\end{itemize}

% In this framework, agents with different abilites would collaborate to complete a given recommendation task. During the process of completing tasks, whether intelligent agents cooperate and the effectiveness of cooperation will be trained. 

% 介绍 actor 和 reflector 更进一步地
% To explore the effectiveness of our MAC framework, 
% Additionally, we primarily focus on the reflection-based cooperation mode, assembling an 'actor' agent for task execution and a 'reflector' agent for feedback and reflection.
% In detail, the actor is responsible for completing the main execution process of the task by having a thought, while the reflector provides crucial feedback and reflections on the actor's decisions. Importantly, the reflection process is not mandatory; the decision "whether to cooperate/reflect" is part of the system's dynamic. It is expected that this actor-reflector collaboration will significantly enhance the accuracy and robustness of our system across various recommendation tasks.
% The reflection paradigm compels LLMs to recognize and amend their prior errors. This extra reflection step enhances the precision and robustness of the model outputs, obliging the recommending agent to reach a more thorough and accurate prediction.

%% file: Sections/2.related_work.tex
\section{Related Work}
% 换一下顺序
\subsection{Agents-based Recommendation}

Currently, research on integrating LLM-based agents for recommendation can be categorized into two primary orientations: \textsl{simulation-oriented} and \textsl{recommender-oriented} approaches. Table~\ref{tab:related_work} compares our MACRec and previous agents-based work.
% 放在2.2后面
%our MACRec in comparison with previous agents-based work.

The \textsl{simulation-oriented} work focuses on using agents to simulate user behaviors and item characteristics in RSs. RecAgent~\cite{zhang2023generative} and Agent4Rec~\cite{wang2023RecAgent} both propose to use agents as user simulators to empower the evaluation of RSs, which feature single-type agents (as users).
AgentCF ~\cite{zhang2023agentcf} explores the simulation of user-item interactions through user-agents and item-agents. It belongs to a multi-type agent system, with only two types and simple interactions.
This line of research aims to provide a deeper understanding of user preferences but falls short of integration into RSs.

The goal of \textsl{recommender-oriented} studies is to build a "recommender agent" with planning and memory components to tackle recommendation tasks. 
InteRecAgent~\cite{huang2023InteRecAgent} and RecMind~\cite{wang2023recmind} primarily focus on improving a single recommender agent's planning and reflection ability.
% 再明确阐述差异 Diverse Rec  Summary: 强调multi-agents的工作都没有开源，缺少demo系统
RAH~\cite{shu2023rah} proposes a human-centered framework using LLM Agents as assistants.
It supports collaboration among different types of agents, yet only in a fixed mode, whereas MACRec enables adaptable collaboration for various uses. Moreover, RAH lacks publicly accessible code or demos.
% 据我们所知，我们是第一个支持xxx,xxx
To the best of our knowledge, MACRec is the first open-source framework supporting multi-type agents for diverse recommendation scenarios.

% to 
% And it is the first open-source framework for multi-agent recommendation.

\subsection{Multi-agent Collaboration}

% The study of multi-agent collaboration is characterized by exploring how multiple intelligent agents interact and collaborate to achieve complex tasks or goals.
% 早期的multi-agent systems
% Early foundations of multi-agent systems originated from distributed artificial intelligence (DAI)~\cite{chaib1992trends} and multi-agent system (MAS)~\cite{stone2000multiagent}. \citet{wooldridge1995intelligent} laid the foundational principles for agent coordination and communication in MAS. 
Multi-agent systems, initially grounded in DAI~\cite{chaib1992trends} and MAS~\cite{stone2000multiagent}, evolved with foundational concepts of agent coordination and communication by \citet{wooldridge1995intelligent}.
% 基于LLM
The advent of powerful LLMs~\cite{brown2020language, gpt42023technical, touvron2023llama, zeng2022glm} has shifted focus towards their application in multi-agent collaboration. ~\citet{brown2020language} demonstrated LLMs' potential in human-like dialogues, applicable to agent-agent communication. ~\citet{vinyals2019grandmaster, nascimento2023gpt} illustrates how LLM agents can collaborate for shared objectives, achieving specific and complex task solutions. 
Recent work~\cite{du2023improving, zhang2023building, chen2023agentverse} leverage multi-agent collaboration to achieve better performance on complex tasks. CAMEL~\cite{li2023camel} and AutoGen~\cite{wu2023autogen} focus on communicative agent systems for complex task solutions through inter-agent dialogue. 
% 加一句总结，没有用在推荐场景
However, existing research on multi-agent collaboration has not investigated its potential in recommendation scenarios.

% ~\citet{madaan2023self}, ~\citet{hao2023reasoning}, ~\citet{shinn2023reflexion} and ~\citet{yao2023retroformer} employ reflection, a framework to optimize the LLM agent corresponding to feedbacks from other agents set as the environment, supplementing LLMs in task adaptation from merely exploiting the knowledge of pretrain model.

%% file: Sections/3.methodology.tex
\begin{figure*}[t!]
\centering
\includegraphics[trim={0 0 0 0}, clip, width=0.9\textwidth]{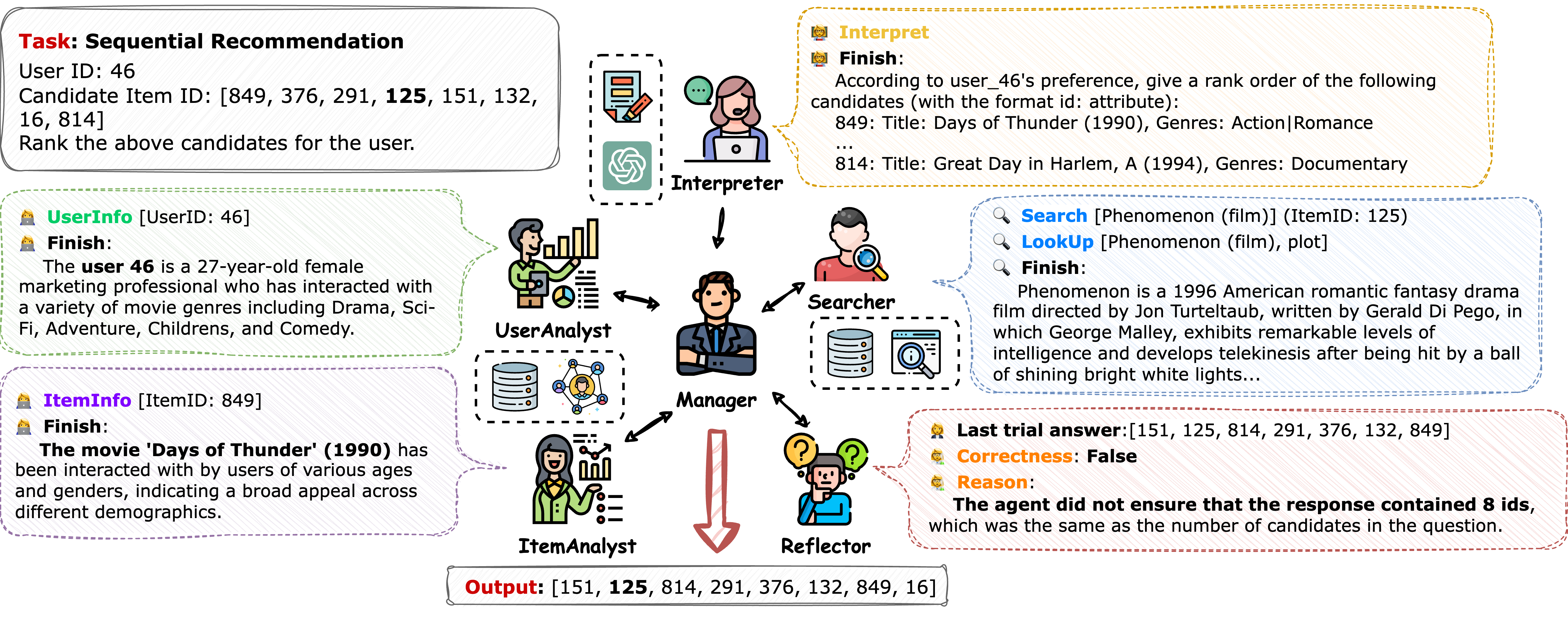}
\caption{The Framework of MACRec. We take a sequential recommendation task as an example to show how these agents work collaboratively.}
\Description[The framework of MACRec.]{The framework of MACRec.}
\label{fig:framework}
\end{figure*}

\section{The MACRec Framework}

\subsection{Framework Overview}

Figure~\ref{fig:framework} illustrates our proposed multi-agent collaboration recommendation framework.
% Framework Overview
% 以图中的sequential recommendation任务为例
A sequential recommendation task is given as an example.

% The \textsl{Manager} is always placed at the center during the running process.
% All other types of agents serve as the assistant role for the \textsl{Manager}.
% 所有的agents以Manager为中心，共同协作完成任务。
% All agents work together to complete the task with the \textsl{Manager} at the center.
% 每个agent有各自的特长，对于给定的任务在完成后会将结果反馈给Manager。
% Each agent has its expertise and will provide feedback on the given sub-task from \textsl{Manager}.
% 一部分Agents支持对工具的调用，如Searcher、Analyst和Task Interpreter。
% In the example provided in Figure~\ref{fig:framework}, 
As shown in the example in Figure~\ref{fig:framework}, 
the \textsl{Task Interpreter} first translates the task in a better way to understand.
Then, as the central component of the entire system, the \textsl{Manager} starts calling other agents to obtain detailed analyses of the user and items.
These agents, including the \textsl{Searcher} and the \textsl{User/Item Analyst}, support the call of some tools, 
% Additionally, some agents support the call of tools, 
% such as the \textsl{Searcher} and the \textsl{User/Item Analyst}.
e.g., the \textsl{Searcher} has access to the search engine and the \textsl{User/Item Analyst} can access detailed information about users and items.
% Manager 1st
After receiving responses from the \textsl{Searcher} and \textsl{Analyst}, the \textsl{Manager} will attempt to provide an answer, i.e., give a ranking order of the candidate sets.
% 解释reflector（每个agent都出现一次）
The \textsl{Reflector} will be responsible for analyzing and reflecting on the \textsl{Manager}'s answer in the last trial and giving suggestions, e.g., modifying the answer format to follow the task requirements.
Eventually, the Manager will reattempt to solve the task based on the reflections and provide a more reasonable answer, e.g., adding the missed item ID.

% 具体每个agent的特点与功能将在下文详细介绍。
The following sections will detail each agent's specific characteristics and functions.\footnote{Code is available at \href{https://github.com/wzf2000/MACRec}{https://github.com/wzf2000/MACRec}.}

% % 我们为不同的推荐场景设计了五类具有各异特长的agents。
% We designed five types of agents with different specialties for different recommendation scenarios.

% \begin{figure}[t!]
% \centering
% \includegraphics[trim={0 0 0 0}, clip, width=0.3\textwidth]{Figures/MACRec-Framework.pdf}
% \caption{The framework of MACRec.}
% \label{fig:framework}
% \end{figure}

% 改一下 summarized， id 46  id 151(Days of Thunder)  (phenomenon) 加上最终输出 加上标号 1 2 3 4 5  题目里写清楚 是一轮的 让底部对齐

\subsection{Agent Roles}

\subsubsection{\textbf{Manager}}
For the given task, the \textsl{Manager} would assign sub-tasks to other agents and complete the main execution process. It oversees the collaboration among all other agents. %, acting like a manager.

% Manager总是交替的执行思考、行动和观察三个步骤。
The \textsl{Manager} always performs the three steps of \textsl{Thought}, \textsl{Action}, and \textsl{Observation} alternately.
% 在思考阶段，Manager会推理当前任务的状态（如分析是否充足，是否需要额外信息等）。
In the \textsl{Thought} phase, the \textsl{Manager} reasons about the current situation of the task (e.g., whether the analysis is sufficient, whether additional information is needed, etc.).
% 而在行动阶段，Manager可选择给出答案结束任务，或是寻求其他agents的帮助。
During the \textsl{Action} phase, the \textsl{Manager} can choose to give an answer to end the task or seek help from other agents (under a particular interface format).
% 其他agents给出的回复将在Manager的observation阶段中给出。
Responses given by other agents will be given in the \textsl{Observation} phase of the \textsl{Manager}.

\subsubsection{\textbf{Reflector}}

The \textsl{Reflector} is responsible for judging the correctness of the answer given by the Manager. A further reflection will be given if the \textsl{Reflector} determines the answer is correct.

% 当Manager即将对同一个任务输入执行第二轮或更多的运行时，Reflector将会介入。
The \textsl{Reflector} will step in when the \textsl{Manager} is about to perform the second or more runs on the same task input.
% 如果Reflector判定Manager给出的答案没有可改进之处，Manager将会不再执行本轮运行，而是延续上一轮的答案。
If the \textsl{Reflector} judges that the answer given by the \textsl{Manager} has no room for improvement, the \textsl{Manager} will no longer perform the current run. % and continue the answer from the previous round.
% 否则，Reflector会进一步总结Manager的可改进之处，如未考虑到用户历史交互序列中的少数高分商品/电影等。
Otherwise, the \textsl{Reflector} will further summarize where the \textsl{Manager} can be improved, e.g., not considering the few highly rated items/movies in the user's historical interactions.

\subsubsection{\textbf{User/Item Analyst}}

\textsl{User/Item Analyst} specializes in examining and understanding the characteristics and preferences of users, as well as the attributes of items.

% Analyst在运行时将会获得所要分析的交互id对，即用户id与物品id。
% The \textsl{Analyst} will get the interaction ID pairs to be analyzed at runtime, i.e. the user ID and item ID.
% Analyst将会获得两类工具的调用权限以辅助分析。
The \textsl{Analyst} will be given access to two tools to assist in the analysis, including info database and interaction retriever.
% 可以通过Info database获得每个用户的用户画像和每个物品的属性信息。
The \textsl{Analyst} can get the user profile of each user and the attributes of each item through the info database.
% 通过Interaction retriever，Analyst可以获得在当前时间之前的用户/物品交互历史。
Through the interaction retriever, the \textsl{Analyst} can get the user/item interaction history before the current time.
% 通过这两种工具的结合，Analyst可以对用户和物品有深入的分析，从而给予Manager准确的建议。
With the combination of these two tools, the \textsl{Analyst} can have an in-depth analysis of the user or the item.

\subsubsection{\textbf{Searcher}}
% The \textsl{Searcher} is tasked with scanning the database or inventory to find items that match the criteria set by the recommendation system.

% Searcher负责将Manager给出的查询需求，配合搜索工具进行检索，最终总结出文本回复给Manager。
The \textsl{Searcher} is responsible for searching under the requirements given by the \textsl{Manager} with the search tool, and finally summarizing the text reply to the \textsl{Manager}.

% 以Wikipedia作为搜索工具为例。
Take Wikipedia as an example of a search tool.
% Searcher可以在通过给出query以获取在Wikipedia中相关度最高的几个词条。
The \textsl{Searcher} can give a search query to get the most relevant entry in Wikipedia.
% Searcher还可以进一步在具体的某个entry中检索存在相关字段的段落。
The \textsl{Searcher} can further retrieve passages in a specific entry where the given keywords exist.
% 通过检索得到结果与Manager给出查询需求的匹配程度，Searcher最终会总结出一段给Manager的文本回复。
Eventually, the \textsl{Searcher} is asked to summarize the paragraph to respond the the \textsl{Manager}'s query.
% Compared the retrieval results with the requirements given by the \textsl{Manager}, the \textsl{Searcher} will eventually summarize a paragraph to respond to the \textsl{Manager}.

\subsubsection{\textbf{Task Interpreter}}
The \textsl{Task Interpreter} translates the dialogs into executable recommendation tasks.

% Task Interpreter将得到用户与系统的对话历史。
The \textsl{Task Interpreter} will get the conversation history when starts running.
% 由于对话历史可能较为冗长，Task Interpreter只会得到历史的最后一部分。
Since conversation histories can be long, the \textsl{Task Interpreter} will only get the last part of the history.
% 同时，Task Interpreter还可以通过文本摘要工具以得到更精炼的历史概括。
The \textsl{Task Interpreter} also has access to call the text summarization tool to get a more concise overview of the history.
% 最终，Task Interpreter会给出一个具体的任务需求描述，用以指导Manager后续的运行。
Eventually, the \textsl{Task Interpreter} will give a specific description of the task requirements that will be used to guide the subsequent runs of the \textsl{Manager}.

%% file: Sections/4.applications.tex
\section{Applications on Recommendation Scenarios}

\begin{figure*}[t!]
% \begin{overpic}[width=1.0\textwidth]{Figures/case.png}
% \put(2,5){\textbf{(a) Interprete the dialog into a task.} }
% \end{overpic}
\centering
\includegraphics[trim={0 0 0 0}, clip, width=0.95\textwidth]{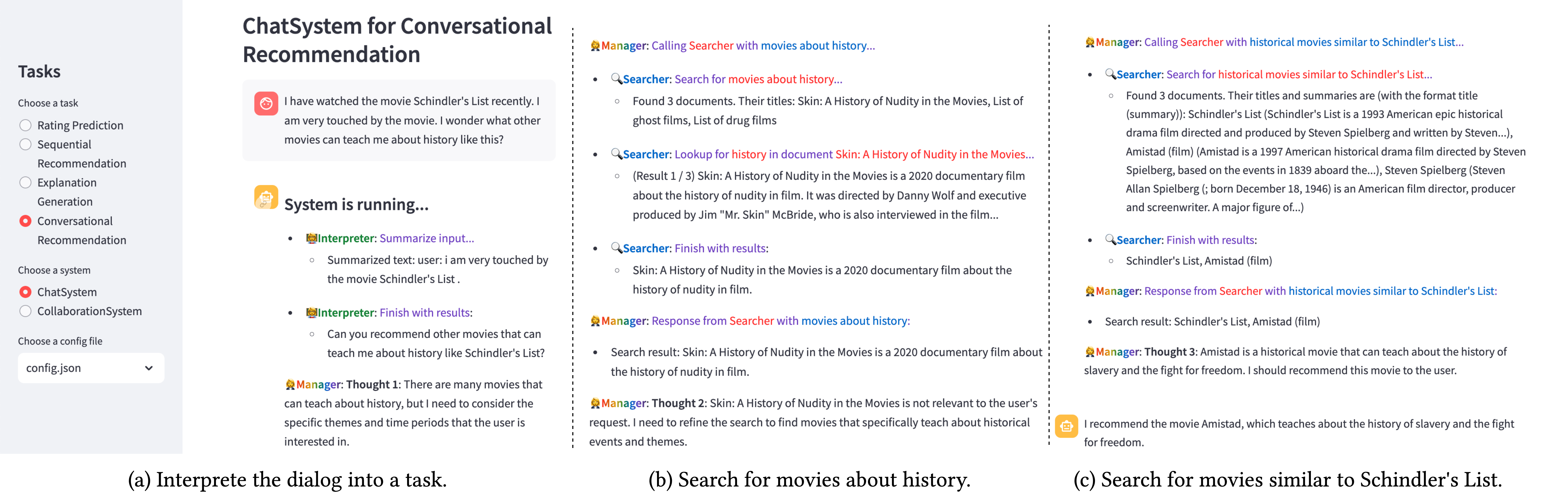}
\caption{The web interfaces of our MACRec, along with a case of how three agents collaboratively address a conversational recommendation task. The interface is composed by the leftmost configuration panel and the main interaction panel.}
\Description[The web interfaces of our MACRec.]{The web interfaces of our MACRec, along with a detailed case study demonstrating the collaborative efforts of three agents in addressing this task. The interface is composed by the leftmost configuration panel and the main interaction panel.}
\label{fig:case}
\end{figure*}
% A usage example of MACRec on a conversational recommendation task.  

Here, we present the applications of MACRec on four recommendation scenarios. Table~\ref{tab:application} summarizes the agents' selection for each scenario.

\begin{table}[h!]
    \centering
    \begin{tabular}{c|ccccc}
    \toprule
    Task & U.Analy. & I.Analy. & Reflector & Searcher & Interpreter \\
    \midrule
    RP & \usym{2611} & \usym{2611} & \usym{2610} \\
    SR & \usym{2611} & \usym{2610} & \usym{2611}\\
    EG & \usym{2611} & \usym{2611} & \usym{2610} & \usym{2611} \\
    CR & & & & \usym{2611} & \usym{2611} \\
    \bottomrule
    \end{tabular}
    \caption{The agents' selection for four applications supported by MACRec.\label{tab:application} \protect\usym{2611} means required and \protect\usym{2610} means optional.}
    \vspace{-3.0em}
\end{table}

\subsection{Rating Prediction (RP)}

% 介绍 Rating Prediction
Rating prediction task involves predicting the numerical rating a user might give to an item, such as a movie or a product, based on their preferences and historical interactions.

% 介绍 MAC
% 在评分预测任务中，每个用户会有不同的打分偏好。
In the rating prediction task, each user will have different rating preferences.
% User Analyst:
The \textsl{User Analyst} can provide a detailed analysis of the user's historical interactions and preferences.
% Item Analyst:
Meanwhile, the \textsl{Manager} also needs characteristic analysis of the target item, which can be provided by the \textsl{Item Analyst}.
% Manager:
% 在两种analyst的帮助下，Manager可以在打分前对于用户的打分倾向和物品的近期评分有具体的了解。
With the help of two types of \textsl{Analysts}, the \textsl{Manager} can know the user's tendency to rate and the item's recent ratings before giving a prediction.

% 我们为评分预测任务使用了Manager、Reflector 和 Analyst（可选的） 三种agent
% We used three agents, the \textsl{Manager}, the \textsl{Reflector}, and the \textsl{Analyst} (optional) for the rating prediction task.
% 我们选用 Reflector 对 Manager 在单轮中可能出现的错误进行识别和归因。
% We chose the \textsl{Reflector} to identify and attribute errors that may occur in a single round of the \textsl{Manager}.
% 同时，Analyst可以帮助更好的总结归纳用户与物品的特征，因此也被我们选取应用在评分预测任务中。
% Meanwhile, the \textsl{Analyst} can help better summarize user and item characteristics, so we also selected it to apply in the rating prediction task.
% For the rating prediction task, the task prompt contains the user's profile, historical interactions, and the target item that needs to be rated.

% yyq: 没有必要介绍prompt
% % 所有的item都会对应给出他们的Attributes，历史交互item还会额外给出用户对其的评分。
% All items will be given their attributes, and items in the user's historical interactions will also be given the user's ratings.
% The actor is asked to rate the target item with the format of \verb|Finish[rating]|, where \verb|rating| is a float number between 1 and 5.

\subsection{Sequential Recommendation (SR)}

% 介绍 Sequential Recommendation
Sequential recommendation systems analyze the sequence of items a user has interacted with to predict their next likely interest.

% 介绍 MAC
% 在序列推荐任务中，对于用户长短期兴趣的建模十分重要。
Modeling of user's long-term and short-term interests is important in sequential recommendation tasks.
Hence, the \textsl{User Analyst}'s role is self-evident.
% 与评分预测任务相比，(由于同时包含了用户的交互历史和候选物品列表)，序列推荐中相关的物品数量明显更多。
The number of relevant items in the sequence is significantly higher than the rating prediction task.
It is hard to ask the \textsl{Item Analyst} to analyze every item that appeared in either the history or the candidate set.
% Therefore, the role of the \textsl{Item Analyst} is comparatively limited compared to other scenarios.
% 而考虑到序列推荐任务的答案更为复杂，Reflector可以很好地帮助Manager发现答案格式上的问题。
Moreover, given that the answers to the sequential recommendation task are much more complex (i.e., a ranking order of the candidate set), the \textsl{Reflector} can help to avoid the \textsl{Manager} getting into formatting troubles.
% 并且单轮的行为分析中，可能遗漏对于用户长期行为的考虑，而对此的反思也是Reflector可以做到的。
A single round of behavioral analysis may omit consideration of long-term user behavior, and reflection on this is something the \textsl{Reflector} can do.

% 序列推荐任务的agents选取与评分预测基本一致。
% The agent selection for the sequence recommendation task is the same as the score prediction.
% 值得注意的是，由于序列推荐任务设定中给出的历史交互物品以及候选集物品的特征较多，由Analyst来负责具体的特征分析可以大大减轻Manager的上下文长度。
% It is worth noting that leveraging the \textsl{Analyst} to do characterization can greatly reduce the context length of the \textsl{Manager}.
% yyq: 没有必要介绍prompt
% For the sequential recommendation task, the task prompt contains the user's profile, historical interactions, and the candidate set that needs to be ranked.
% % 候选集中包含了用户真实的下一个交互item和若干个未交互的负样本。
% The candidate set contains the positive sample (the next item in the user's historical interactions) and several negative samples (items that the user has not interacted with).
% % 相关物品的Attributes也会如上给出。
% The attributes of the relevant items will also be given as above.
% The actor is asked to rank the candidate set with the format of \verb|Finish[rank]|, where \verb|rank| is a permutation of the item IDs of the candidate set.

\subsection{Explanation Generation (EG)}
% 介绍 Explanation Generation
This task involves generating understandable and relevant explanations for the recommendations provided to users.

% 介绍 MAC
The explanation generation task also requires a detailed analysis of both the user and the item.
In addition, more information about the item may also help the \textsl{Manager} understand the user's behavior towards it.
% 例如，用户可能对同一个导演的多部电影拥有相似的喜好，而这些信息不一定被数据集所包含。
For example, a user may have similar preferences for multiple movies by the same director.
The information about the director may not be contained in the dataset.
Retrieving these extra pieces of information is suitable for the \textsl{Searcher} to perform.

% 我们在解释生成任务上的设定与前两个任务基本一致。
% Our setup for the explanation generation task is similar to the first two tasks.
% 除了Reflector和Analyst以外，我们还选取了Searcher以满足解释生成对于具体物品（如电影）的进一步深入了解的需求。
% In addition to the \textsl{Reflector} and the \textsl{Analyst}, we selected the \textsl{Searcher} (optional) to satisfy the need for further insights into specific items (e.g., movies) for explanation generation.

\subsection{Conversational Recommendation (CR)}
% 介绍 Conversational Recommendation
Conversational recommender systems engage users in a dialogue to refine their preferences and deliver more accurate suggestions.

% 介绍 MAC
In conversational scenarios, the user's input text is not necessarily explicitly instructive.
Hence, the \textsl{Task Interpreter} can help translate the conversation history into a more concise and clear task prompt.
% 同时，用户的输入需求中可能包含Manager所不了解的物品信息。
In addition, the user's input requirements may contain information unknown to the \textsl{Manager}. % , e.g., a product that has not sold on the platform.
In this case, the \textsl{Searcher} can help the Manager understand what the user mentioned.

% Beyond the above applications, TODO
% 还可以支持其他的task，需要XXX可以支持
Beyond the abovementioned applications, our framework can support other scenarios by customizing the configurations.

%% file: Sections/5.case_example.tex
\section{Interface Demonstration}

% (a)往右挪 面板下面加解释，或者图中加解释
Figure~\ref{fig:case} presents the web interfaces of our framework, along with a detailed case study demonstrating the collaborative efforts of three agents in addressing a conversational recommendation task. 

% User Interface of MACRec
The interface can be divided into two main panels. 
1) \textbf{Configuration panel}, where users can select different tasks to tackle, such as "Rating Prediction." Users can also customize different systems and configuration files for the task execution. 
2) \textbf{Interaction panel}, where the whole collaboration process takes place. Agents with different abilities would complete the task collaboratively.
% Agents with different abilities would 

% Case on Conversational Recommendation
In Figure~\ref{fig:case}, the user has expressed a preference for the movie "Schindler's List" and seeks recommendations for similar historical movies. 
% As shown in subfigure (a),
The \textsl{Interpreter} summarizes this input and translates it into a clearer task.
Then, the \textsl{Manager} calls for the help of the \textsl{Searcher} for two rounds, searching for movies about history and movies similar to "Schindler's List".
According to all the information, the \textsl{Manager} gives the final recommendation movie "Amistad".

%% file: Sections/6.conclusion.tex
\section{Conclusion}
% 可以缩减篇幅 大概8行就行 强调table 1中的5个维度
In this work, we propose a novel LLM-based multi-agent collaboration framework for recommendation, called MACRec.
Unlike existing studies on using agents for user/item simulation, we directly tackle recommendation tasks through the collaboration of various agents. 
% , including managers, user/item analysts, reflectors, interpreters, etc.
We present applications of MACRec on four different recommendation tasks. %, including rating prediction, sequential recommendation, conversational recommendation, and explanation generation of recommendation results. 
Moreover, We developed an online web interface for MACRec, visualizing how agents work collaboratively. 
% The case study on a conversational recommendation task primarily demonstrates the effectiveness of our framework.
% Preliminary experiments have demonstrated the feasibility and effectiveness of our framework.

% Future work TODO

% We also design a multi-round off-line training to obtain agents equipped with recommendation knowledge. With tasks including rating prediction and sequential recommendation, we have demonstrated the feasibility and effectiveness of our framework.

%% file: Appendix/appendix.tex
% \section{Prompt Template}

%% file: Appendix/acknowledgment.tex
% \section{Acknowledgments}